\documentclass[twocolumn,prd,groupedaddress,nofootinbib,showpacs]
{revtex4}

\usepackage{latexsym}
\usepackage{amssymb}
\usepackage{graphicx}
\usepackage{dcolumn}
\usepackage{bm}

\begin{document}

\title{ Nonequivalent Seiberg-Witten maps for noncommutative massive U(N) 
gauge theory  }
\author{Ricardo Amorim$^{1,a}$, Nelson R. F. Braga$^{1,b}$ and  
Cristine N. Ferreira $^{1,2,c}$}

\affiliation{$^1$Instituto de F\'{\i}sica, 
Universidade Federal do Rio de Janeiro,
Caixa Postal 68528, RJ 21941-972 -- Brazil
\\
$^2$Grupo de F\'{\i}sica Te\'orica Jos\'e Leite Lopes (GFT),
Petr\'opolis, RJ, Brazil }

\date{\today}

\begin{abstract}
Massive vector fields can  be described in a gauge invariant way 
with the introduction of compensating fields. In the unitary  gauge one recovers
the original formulation.
Although this gauging mechanism can be extended to noncommutative spaces in 
a straightforward way,  non trivial aspects show up when we consider the 
Seiberg-Witten map. As we show here, only a particular class of its solutions 
leads to an
action that   admits the unitary gauge fixing.

\end{abstract}

\maketitle

\section{Introduction}

\renewcommand{\theequation}{1.\arabic{equation}}
\setcounter{equation}{0}

The idea that spacetime may be noncommutative at very small length scales
is not new\cite{Snyder}. 
Originally this has been thought just as a mechanism for providing 
space with a natural cut off that would control ultraviolet divergences.
However, the interest on this topic  increased a lot in the last years 
motivated mainly by important results  coming from string theory that 
indicate a possible noncommutative structure for space time 
(see \cite{SW},\cite{REVIEW} for a review and a wide list of 
important references). 
The presence of an antisymmetric tensor background along the D-brane\cite{Po} 
world volumes (space time region where the string endpoints are located) 
is an important source for noncommutativity  in string theory\cite{HV,CH1}.
  
In noncommutative space-time of dimension $D$ the coordinates  $x^\mu$ are  
replaced by Hermitian generators $\hat x^\mu$ of a noncommutative $C^*$-algebra 
over spacetime functions satisfying

\begin{equation}
[\hat x^\mu,\hat x^\nu]=i\theta^{\mu\nu}
\end{equation}

\noindent where $\theta^{\mu\nu}$ is usually taken as   a
constant antisymmetric  matrix of dimension   $D$. 

In order to define noncommutative quantum field theories one can 
rather than working with  noncommuting functions of the operators 
$\hat x^\mu$, replace the ordinary products everywhere by the  Moyal star product

\begin{equation}
\phi _{1}(x)\star \phi _{2}(x)=\exp \,\left( \frac{i}{2}
\theta ^{\mu \nu}\partial _{\mu }^{x}\partial _{\nu }^{y}\right) \,
\phi _{1}(x)\phi_{2}(y)|_{x=y}
\label{01}
\end{equation}

\noindent and then consider usual functions of
$x^\mu$. 
Since the space time integral of the Moyal product of two fields is equal 
to the usual product (when boundary terms do not contribute), the noncommutativity 
does not affect the free part of the action but the vertices. 
This  implies many interesting features of
noncommutative quantum field theories as discussed in \cite{REVIEW,SW}.

Gauge theories can be extended to noncommutative spaces by considering actions 
that are invariant under gauge transformations defined in terms of the Moyal structure. 
However, the form of these gauge transformations imply that the algebra  of the 
generators must be closed not only under commutation  but also under 
anticommutation. So $U(N)$ is usually chosen as the symmetry group for
noncommutative extensions of Yang-Mills theories in place of $SU(N)$, 
although other symmetry structures can also be 
considered \cite{Bonora}\cite{Wess}\cite{AF}.

Once one has a noncommutative gauge theory, in the sense that the field
polinomia in the action and their gauge structure are constructed by using
 Moyal products, it is possible to generate a map from this noncommutative
theory to an ordinary one, as shown by Seiberg and Witten \cite{SW}. 
Interesting aspects of the general form of this map can be found in \cite{AK}.
The mapped  Lagrangian is usually  written as a nonlocal infinite series of 
ordinary fields and their space-time derivatives but the noncommutative 
Noether identities are however kept by the Seiberg-Witten map. 
\medskip

It is sometimes useful to transform global symmetries in gauge symmetries 
by the introduction of pure gauge ``compensating fields" \cite{CF}. 
This procedure can be used, for example, as a tool for calculating anomalous 
divergencies associated with global currents\cite{ABH}.
Another use of compensating fields is to allow  a gauge invariant 
formulation for a massive vector field. 
In this letter we will investigate the extension to noncommutative spaces of this kind 
of gauging process.
We will see that it is possible to define a noncommutative version of
a gauged vector field with mass  and also that a Seiberg-Witten map can be constructed.
When we introduce a gauge invariance that was not originally
present it is in general possible to return to the original theory by a particular 
gauge fixing of this new symmetry. 
This condition, expected to  hold also at non commutative level, 
will represent a criterion for choosing the appropriate 
Seiberg-Witten map among the general solutions.

\medskip

This paper is organized as follows: in section {\bf II}  we discuss the 
noncommutative massive vector field theory. In section {\bf III} we present the general
structure of the Seiberg-Witten map, that means: we derive the general  set of equations
it has to satisfy.  Different solutions for the map are then presented in section 
{\bf IV}. We reserve section {\bf V} for some concluding remarks.

\section{Gauging  the noncommutative $U(N)$ Proca field}
\renewcommand{\theequation}{2.\arabic{equation}}
\setcounter{equation}{0}

The  action for the ordinary  $U(N)$ Proca (massive vector) field is given by

\begin{equation}
\label{Action}
S[a]=\,tr\int d^{4}x\,\left(\,-\frac{1}{2}f_{\mu \nu } f^{\mu \nu } 
+m^2 a_\mu a^\mu\right)
 \label{2.1}
\end{equation}

\noindent where the curvature tensor is defined by

\begin{equation}
f_{\mu \nu } =\partial _{\mu }a_{\nu }-\partial _{\nu }a_{\mu
}-i\,[a_{\mu } , a_{\nu }]  \label{2.2}
\end{equation}

\noindent and the vector field $a_\mu$ take values in the $U(N)$ algebra, with
generators $T^{A}$, assumed to be normalized as

\begin{equation}
tr(T^{A}T^{B})={\frac{1}{2}}\delta ^{AB}  \label{2.3}
\end{equation}

\noindent and satisfying the (anti)commutation relations

\begin{eqnarray}
\lbrack T^{A},T^{B}] &=&if^{ABC}T^{C}  \nonumber \\
\{T^{A},T^{B}\} &=&d^{ABC}T^{C}  \label{2.4}
\end{eqnarray}

We take $f^{ABC}$ and $d^{ABC}$ as
completely antisymmetric and completely symmetric respectively.
\medskip

The theory described by (\ref{2.1}) is not gauge invariant because of the presence of  
the mass term.  
As it is well known, it is possible to gauge the above theory with the introduction
of compensating fields. In the Lagrangian formalism, this can be directly done with the
introduction of  scalar fields $g$ which transform as $U(N)$ group elements. 
The procedure is very simple and consists in replacing the field $a_\mu$ by a kind of 
invariant collective field $\tilde a_\mu=\tilde a_\mu( a, g)$ 
defined as\cite{CF,ABH}

\begin{eqnarray}
\tilde a _{\mu }&=&g^{-1}\,a_\mu\,g+i\,g^{-1}\partial_\mu g\nonumber\\
&=&g^{-1}\,(a_\mu-b_\mu)\,g
\label{2.5}
\end{eqnarray}

\noindent where

\begin{equation}
b_\mu=-i\,\partial_\mu g \,g^{-1}
 \label{2.6}
\end{equation}

\noindent is a "pure gauge" compensating vector field since its curvature, 
constructed as in (\ref{2.2}), vanishes identically. As $a_\mu$, $b_\mu$ also
takes values in the $U(N)$ algebra.
 
If we write $\tilde a_\mu$ instead of $a_\mu$ in action
(\ref{2.1}), we get directly 

\begin{equation}
\label{Action2}
S[a,g]= tr\int d^{4}x \left(\,-\frac{1}{2}f_{\mu \nu } f^{\mu \nu } +
m^2 (a_\mu-b_\mu)( a^\mu - b^\mu)\right)
 \label{2.7}
\end{equation}

\noindent which is now invariant under the gauge transformations

\begin{eqnarray}
\bar\delta a _{\mu }&=&
\partial_\mu \alpha\ - i[a_{\mu},\alpha ] \,\equiv \,D_\mu \alpha
\nonumber\\
\bar\delta g&=&i\alpha\,g
\label{2.8}
\end{eqnarray}

\noindent as can be verified. We are denoting the gauge variation by $\bar\delta$
since we will reserve the symbol  $\delta$ for the gauge variation of the 
noncommutative case, which will be shortly introduced. For completeness, 
we note that the above definitions imply that

\begin{equation}
\bar\delta b_\mu\,=\, \partial_\mu\alpha - i[b_\mu , \alpha]
\,\equiv\,\bar D_\mu\alpha
\label{2.9}
\end{equation}

\noindent The gauge algebra of all of these fields closes as

\begin{equation}
[\bar\delta_1,\bar\delta_2]\,y=\bar\delta_3\, y
\label{2.10}
\end{equation}

\noindent $y$ representing  $a_\mu$, $g$ or $b_\mu$.
The parameter composition rule then is given by

\begin{equation}
\alpha_3 = i[\alpha_2,\alpha_1]
\label{2.11}
\end{equation}

As expected, the original theory is recovered in the unitary gauge $g=1$.
There is no  obstruction to implement this model also at the quantum level,
even if there are arbitrary couplings with fermions \cite{CF}, since
candidates to anomalies are compensated by appropriate Wess-Zumino terms
constructed with the fields $a_\mu$ and $g$.
\medskip

The gauge invariant action given by (\ref{2.7}) can be extended to a
noncommutative space. 
Let us represent the corresponding noncommutative fields by capital letters
and introduce Moyal products whenever usual ordinary products appear in 
the original ordinary theory. We get the non commutative version for the
action (\ref{Action2})

\begin{equation}
S=\,tr\int d^{4}x\,\left(\,-\frac{1}{2}F_{\mu \nu } F^{\mu \nu } +
m^2 (A_\mu-B_\mu)( A^\mu - B^\mu)\right)
 \label{2.12}
\end{equation}

\noindent where now the curvature is given by

\begin{equation}
F_{\mu \nu } =\partial _{\mu }A_{\nu }-\partial _{\nu }A_{\mu}-i\,[A_{\mu } 
\buildrel\star\over, A_{\nu }] 
\label{2.13}
\end{equation}

\noindent and the infinitesimal gauge transformations (\ref{2.8}) are replaced by

\begin{eqnarray}
\delta A_{\mu }&=&D_\mu \epsilon\nonumber\\
&=&\partial_\mu \epsilon\ - i[A_{\mu} \buildrel\star\over, \epsilon]\nonumber\\
\delta F_{\mu \nu} &=& -i[ F_{\mu \nu}\buildrel\star\over,\epsilon] \nonumber\\ 
\delta G&=&i\epsilon \star G
\label{2.14}
\end{eqnarray}

\noindent 
Note that we are using the same symbol to denote ordinary and noncommutative
covariant derivatives but we believe that there will be no misunderstanding.
The compensating field $B_\mu$ is now

\begin{equation}
B_\mu\equiv -i\,\partial_\mu G \star G^{-1}
\label{2.15}
\end{equation}

\noindent and transforms accordingly

\begin{eqnarray}
\delta B_{\mu }&=&\bar D_\mu \epsilon\nonumber\\
&=&\partial_\mu \epsilon\ - i[ B_\mu \buildrel\star\over,\epsilon]\,\,.
\label{2.16}
\end{eqnarray}

\noindent Its noncommutative curvature, defined in analogy with 
(\ref{2.13}), vanishes identically as in the ordinary case. 
As expect, the noncommutative gauge transformations listed above also
close in an algebra:

\begin{equation}
[\delta_1,\delta_2]\,Y=\delta_3\, Y
\label{2.17}
\end{equation}

\noindent  $Y$ representing $A_\mu$, $G$ or $B_\mu$.
The composition rule for the parameters now is given by

\begin{equation}
\epsilon_3=i[\epsilon_2\buildrel\star\over,\epsilon_1]
\label{2.18}
\end{equation}

\noindent and belongs to the algebra due to (\ref{2.4}).
In the above expressions $G$ is an element of the noncommutative $U(N)$ group. 
This means that the composition rule is also to be operated  with the Moyal product. 
For instance the inverse to $G$ is defined by $G^{-1}\star \,G=1$ which 
implies different features  when compared with the usual ( commutative ) $U(N)$ group. 
If one writes down explicitly expressions like (\ref{2.15}), (\ref{2.16})
or (\ref{2.18}),
it is easy to see that they will involve both the structure functions 
$f^{ABC}\,$ and $\,d^{ABC}\,$ present in eq. (\ref{2.4}). 
With these remarks in mind, we see that there is also 
no problem for implementing the unitary gauge $G=1$. This can be seen by using directly 
the finite form of the gauge transformations (\ref{2.14}):

\begin{eqnarray}
 A^\prime_{\mu }&=&U^{-1}\star A_\mu\star U+i U^{-1}\star \partial_\mu U\nonumber\\
 G^\prime&=&iU^{-1}\star G
\label{2.171}
\end{eqnarray}

\noindent This guarantees that the physical content of the Proca model is not 
affected by the introduction of the compensating fields.
We observe that the Hamiltonian treatment of these points has been done for 
the simpler noncommutative $U(1)$ case  \cite{AB}, along the  BFFT procedure \cite{BFFT}.

\bigskip

\section{General structure of the Seiberg-Witten map}
\renewcommand{\theequation}{3.\arabic{equation}}
\setcounter{equation}{0}

Let us consider now the Seiberg-Witten map linking the massive noncommutative
$U(N)$ gauge theory described in the previous section and a corresponding 
higher derivative theory defined in terms of usual commutative products and 
ordinary fields. Following the same notation employed in the last section, the 
noncommutative variables will be represented by capital letters, here 
generically denoted by $Y$. The corresponding ordinary ones, represented by 
small letters, will be generically   denoted by 
$y$. We assume that  the gauge transformations $\delta Y$ of the noncommutative 
variables listed in the last section can be obtained through the underlying gauge 
structure of the corresponding ordinary theory. The construction of the Seiberg-Witten 
map starts by imposing for all  fields that 

\begin{equation}
\delta Y=\bar\delta Y[y]
\label{SW1}
\end{equation}

The explicit form of this map  comes solving the above equations when one
assumes that the noncommutative parameters $\epsilon$ are functions
of the commutative parameters $\alpha$ and ordinary fields $y$. Although we are
taking the same form of the gauge transformations displayed in (\ref{2.8})
and (\ref{2.14}), the form of the mapped action will be different from (\ref{2.7})
if the map is nontrivial. Now, the transformations above also close in an algebra:

\begin{widetext}

\begin{eqnarray}
& &\left[\,\bar\delta_1,\bar\delta_2\,\right]\,  A_\mu [y]=D_\mu\left(
\,\bar\delta_1\epsilon_2[y]-
\bar\delta_2\epsilon_1[y]+
i\left[\epsilon_2[y]\buildrel\star\over,\epsilon_1[y]\right]\,\right)
\nonumber\\
& &\phantom{\left[\,\bar\delta_1,\bar\delta_2\,\right]\, \Pi_i[y]\,}=
D_\mu\epsilon_3[y]
\nonumber\\
& &\,\,\,\left[\,\bar\delta_1,\bar\delta_2\,\right]\,G[y]=i\left(
\,\bar\delta_1\epsilon_2[y]-
\bar\delta_2\epsilon_1[y]+
i\left[\epsilon_2[y]\buildrel\star\over,\epsilon_1[y]\right]\,\right)G[y]\nonumber\\
& &\phantom{\left[\,\bar\delta_1,\bar\delta_2\,\right]\, \Pi_i[y]\,}=i\epsilon_3[y]G[y]
\label{SW2}
\end{eqnarray}

\end{widetext}

\noindent where the indices $1$, $2$ and $3$ represent the dependence of 
$\epsilon$ in $\alpha_1$, $\alpha_2$ and $\alpha_3$. For instance, 
$\epsilon_3[y]\equiv\epsilon[\alpha_3,y]$.
From the equations above we find the composition
rule for the noncommutative parameter $\epsilon[y]$ given by

\begin{equation}
\label{SW3}
\epsilon_3[y]=
\bar\delta_1\epsilon_2[y]-
\bar\delta_2\epsilon_1[y]+
i\left[\epsilon_2[y]\buildrel\star\over,\epsilon_1[y]\right]
\end{equation}

\noindent in place of (\ref{2.18}). 
Equation (\ref{SW2}) is not new in the literature \cite{SW}\cite{Wess} but will be crucial 
for the results that we will derive.

Now let us obtain the general equations that must be satisfied by the Seiberg-Witten map.
Assuming, as usual, that the gauge transformation parameter 
can be expanded to first order in $\theta^{\mu\nu}$ as
 $\epsilon[y]=\alpha+\epsilon^{(1)}[y]$, 
we get from (\ref{SW3}) that

\begin{eqnarray}
\bar\delta_1\epsilon^{(1)}_2-\bar\delta_2\epsilon^{(1)}_1
&-&i[\alpha_1,\epsilon^{(1)}_2]+i[\alpha_2,\epsilon^{(1)}_1]
- \epsilon^{(1)}_3 \nonumber\\
&=&-\frac{1}{2}\theta^{\mu\nu}\{\partial_\mu\alpha_1,\partial_\nu\alpha_2\}
\label{SW4}
\end{eqnarray}

\noindent This relation will be important in finding the Seiberg-Witten map for the
gauge parameter. We will see in the next section that it allows more than one solution
for $\epsilon^{(1)}$.
Assuming as well that to first order in  $\theta$ the field is expanded as
 $A_\mu=a_\mu + A^{(1)}_\mu$, the 
field strength $F_{\mu \nu} = f_{\mu \nu} + F_{\mu \nu}^{(1)}$ and that $G=g+G^{(1)}$, 
it is not difficult to deduce from (\ref{2.8}), (\ref{2.14}) and (\ref{SW1}) that

\begin{equation}
\bar\delta A^{(1)}_\mu+i[A^{(1)}_\mu, \alpha ]=\partial_\mu\epsilon^{(1)}+
i[\epsilon^{(1)},a_\mu]
-\frac{1}{2}\theta^{\alpha\beta}\{\partial_\alpha\,\alpha,\partial_\beta a_\mu\}
\label{SW5}
\end{equation}

\noindent and as a consequence, the field strength transformation satisfy

\begin{equation}
\bar\delta F^{(1)}_{\mu \nu} + i[F^{(1)}_{\mu \nu}, \alpha ] = i[\epsilon^{(1)},
f_{\mu \nu}]
-\frac{1}{2}\theta^{\alpha\beta}\{\partial_\alpha\,\alpha,\partial_\beta f_{\mu \nu}\}
\label{SW51}\,\,.
\end{equation}

Also from the same equations we get 
 
\begin{equation}
\bar\delta G^{(1)} - i\alpha G^{(1)}= - \frac{1}{2}\theta^{\mu\nu}\partial_\mu\,
\alpha\partial_\nu\,g + i\epsilon^{(1)}\,g
\label{SW6}
\end{equation}

\noindent for the compensating field $G$. 
The corresponding vector field, writing in first order in 
$\theta $  that $B_{\mu} = b_{\mu} + B^{(1)}_{\mu}$, satisfies

\begin{equation}
\bar\delta B^{(1)}_\mu+i[B^{(1)}_\mu , \alpha]=\partial_\mu\epsilon^{(1)}+
i[\epsilon^{(1)},b_\mu]
-\frac{1}{2}\theta^{\alpha\beta}\{\partial_\alpha\,\alpha,\partial_\beta b_\mu\}
\label{SW61}
\end{equation}

Instead of solving the above equation, we observe that
the map for $G$ induces directly a map for $B_\mu$. From (\ref{2.15})  
one can show that

\begin{eqnarray}  
B^{(1)}_\mu &=& -i\partial_\mu g (G^{-1})^{(1)}-i\partial_\mu G^{(1)} (g^{-1})
\nonumber\\
&+& \frac{1}{2}\theta^{\alpha\beta}
\partial_\alpha\partial_\mu\,g\partial_\beta(g^{-1}) 
\label{SW11}
\end{eqnarray}

\noindent solves (\ref{SW61}). 
Now, by using the equations for the gauge transformations defined above,
it is not difficult to verify that action  (\ref{2.12}) written as


\begin{eqnarray}
S &=& tr\int d^4x\Big( -\frac{1}{2}f_{\mu \nu} f^{\mu \nu} + m^2(a_{\mu} - 
b_{\mu})(a^{\mu} - b^{\mu})\nonumber\\
&-& f^{\mu \nu}F_{\mu \nu}^{(1)} +  m^2\{a^{\mu} -
b^{\mu}, A_\mu^{(1)} - B_\mu^{(1)} \} \Big) \label{3.11}
\end{eqnarray}


\noindent up to $O(\theta^2)$, is indeed gauge invariant. This result is of course independent 
of the particular maps one obtains from (\ref{SW5}), (\ref{SW6}) or
 (\ref{SW61}). 

\section{Different solutions of Seiberg-Witten map} 
\renewcommand{\theequation}{4.\arabic{equation}}
\setcounter{equation}{0}

Let us now look for the solutions  of the Seiberg-Witten map.
The general  solution of (\ref{SW4}) when the compensating field sector is not present 
is\cite{AK} 

\begin{equation}
\epsilon^{(1)}=\frac{1}{4}\theta^{\mu\nu}\left\{\partial_\mu\alpha,a_\nu\right\}
\,+\,\lambda_1 \theta^{\mu\nu}\left[\partial_\mu\alpha,a_\nu\right]\,.
\label{SW7}
\end{equation}

\noindent where $\lambda_1$ is an arbitrary constant.
 The first term corresponds to the particular solution of eq.(\ref{SW4})
and the second term is the solution of the homogeneous part of the same equation.
It is possible 
from (\ref{SW5}) and (\ref{SW7}) to find an explicit form for the map of the 
connection as\cite{AK} 

\begin{eqnarray}
A_\mu[a]&=& a_\mu-\frac{1}{4}\theta^{\alpha\beta}\left\{a_\alpha,\partial_\beta a_\mu
+f_{\beta\mu}\right\}\,+ 
\sigma\theta^{\alpha\beta}D_\mu f_{\alpha\beta}\nonumber \\
&+& {\lambda_1\over 2} \theta^{\alpha\beta}
{ D}_\mu [ a_\alpha , a_\beta ]
+ O(\theta^2)
\,.
\label{SW9}
\end{eqnarray}

\noindent where $\sigma$ is also an arbitrary constant associated with the 
homogeneous solution of (\ref{SW5}) when one uses (\ref{SW7}). 
We observe that if we consider only the particular solution 
($\lambda_1=0$) for the
gauge parameter, eqs. (\ref{SW6}) and (\ref{SW7}) give us

\begin{eqnarray}  
G[a,g]&=&g-\frac{1}{2}\theta^{\alpha\beta}a_\alpha\left(\partial_\beta 
g-\frac{i}{2}a_\beta g\right)\nonumber\\
&+&\gamma \theta^{\alpha\beta} f_{\alpha\beta}g +O(\theta^2)
\label{SW10}
\end{eqnarray}

\noindent where $\gamma$ is arbitrary. At this point we note that it if we choose the ordinary unitary gauge $g = 1$
the corresponding non commutative mapped group element keeps a dependence on 
$a_\mu$ and can not be suppressed from the theory as can be seen from the above 
expression.  However by considering the complete solution (\ref{SW7}) and taking $\lambda_1 = -1/4$ 
it is possible to eliminate one of the problematic  terms in (\ref{SW10}) to obtain

\begin{equation}  
G[a,g]=g-\frac{1}{2}\theta^{\alpha\beta}a_\alpha \partial_\beta 
g  + \gamma \theta^{\alpha\beta} f_{\alpha\beta}g +O(\theta^2)
\label{SW10a}
\end{equation}

If we now choose $\gamma=0$, $G$ goes to $g$ in the unitary gauge. Also, from 
(\ref{SW11}), 

\begin{equation}  
B^{(1)}_\mu=\frac{1}{2}\theta^{\alpha\beta}\left((\bar D_\mu b_\alpha)b_\beta 
-\bar D_\mu(a_\alpha b_\beta)\right) 
\label{SW12}
\end{equation}

\noindent when one uses (\ref{SW10a}) with $\gamma=0$.
Observe, however, that the expression for $A_\mu^{(1)}$ coming from (\ref{SW9}),  
with $\lambda_1 = -1/4$ does not vanish for any $\sigma$. We will show in what 
follows that when we consider the $g$ sector, it is possible
to construct a Seiberg-Witten map that can be completely suppressed in the unitary gauge.
We are considering a theory involving the pure gauge field $b_\mu$ besides the
usual gauge field $a_\mu$. 
So, the space of  solutions for $\epsilon^{(1)},G^{(1)},A_\mu^{(1)}, B_\mu^{(1)}$
representing the non commutative field extensions is actually greater than 
the one studied in detail in \cite{AK}. 
One can check that now instead of (\ref{SW7}) we get 

\begin{eqnarray}
\epsilon^{(1)}&=&\frac{1}{4}(1 - \rho )\theta^{\mu \nu }\left\{\partial_\mu\alpha,
a_\nu\right\} + 
\lambda_1\theta^{\mu \nu }\left[\partial_\mu\alpha,
a_\nu\right]\nonumber\\
& +& \frac{1}{4}\rho \theta^{\mu \nu }\left\{\partial_\mu\alpha,
b_\nu\right\}
+ \lambda_2\theta^{\mu \nu }\left[\partial_\mu\alpha,
b_\nu\right]
\label{SW13}
\end{eqnarray}

\noindent when one also considers the compensating field sector. 
Observe that the first and third terms play a complementary role as a 
particular solution of eq. (\ref{SW4}). The other terms represent homogeneous 
solutions. In  Eq. (\ref{SW13}) $\rho , \lambda_1 $ and $\lambda_2$ are arbitrary.

From (\ref{SW6})  and (\ref{SW13}) we get now 

\begin{eqnarray} 
G[a,g]&=&g-\frac{1}{2}(1-\rho)\theta^{\alpha\beta}
a_\alpha\left(\partial_\beta g-\frac{i}{2}a_\beta g\right)\nonumber\\
&+&i\lambda_1\theta^{\alpha\beta}a_\alpha a_\beta g
+\gamma \theta^{\alpha\beta} f_{\alpha\beta}g \nonumber\\
&+&
i(\lambda_2-\frac{\rho}{4})\theta^{\alpha\beta}b_\alpha b_\beta g
+
O(\theta^2)
\label{SW14}
\end{eqnarray}

\noindent 
Since $b_\alpha$ vanishes identically when $g$ goes to $1$, it is possible 
to implement an unitary gauge  for $G(g)$ 
if we choose $\lambda_1=\frac{1}{4}(\rho-1)$ and $\gamma=0$, 
leaving $\lambda_2$ free. This choice, however, does not
make $A^{(1)}\rightarrow 0$ when $g\rightarrow 1$, 
as can be observed from (\ref{SW5}) and (\ref{SW13}). 
Additionally imposing  that  $\rho=1$ and
$\lambda_2=0$, we verify that $A^{(1)}\rightarrow 0$ when $g\rightarrow 1$.
In this last case

\begin{equation}
B^{(1)}_\mu=
\frac{1}{4}\theta^{\alpha\beta}\{\,{\bar D}_\mu b_\alpha,b_\beta\,\} 
=\frac{1}{4}\theta^{\alpha\beta}\{\,\partial_\alpha b_\mu , b_\beta\,\} 
\label{SW15}
\end{equation}

\noindent and

\begin{equation}        
A^{(1)}_\mu=\frac{1}{4}\theta^{\alpha\beta}\left\{b_\alpha,  D_\mu b_\beta - 
2 \partial_\beta a_\mu\right\}
\label{SW16}
\end{equation}

\noindent and indeed both expressions  vanish in the unitary gauge. 
This is in accordance with the fact that the original Proca model is not a gauge theory.
\medskip

Now that the  structure of this map has been found, it is  only algebraic 
work the construction of the  corresponding mapped action.
From (\ref{2.13}) 

\begin{eqnarray}
F_{\mu \nu} &=& f_{\mu \nu} + D_\mu A_\nu^{(1)}-D_\nu A_\mu^{(1)}
+\frac{1}{2}\theta^{\alpha\beta}
\left\{\partial_\alpha a_\mu,\partial_\beta a_\nu\right\}\nonumber\\
&\equiv&f_{\mu \nu}+F^{(1)}_{\mu \nu}
\label{SW17}
\end{eqnarray}

\noindent up to $O (\theta^2 )$, and discarding terms that come 
from the homogeneous part of (\ref{SW51}) \cite{AK} that do not vanish if $g=1$. 
Now, the mapped action can be written as in (\ref{3.11}) with 
$ B^{(1)}_\mu , A^{(1)}_\mu $ and $ F^{(1)}_{\mu \nu}$ given by 
(\ref{SW15})-(\ref{SW17}).

This action is invariant under the  transformations 
(\ref{2.8}) and (\ref{2.9}) since condition (\ref{SW1}) defining the 
Seiberg-Witten map is satisfied by construction. This guarantees that the 
Noether identities are kept by the map. 
Also, the unitary gauge: $g=1$, $b_\mu \,=\,0$ can be implemented in a consistent
way recovering the non commutative Proca model action given by (\ref{2.12}), 
with $B_\mu = 0$ and $A_\mu=a_\mu$, in $O(\theta^2)$.

\section{Conclusion}

We discussed here how to build up a noncommutative extension
for a gauged massive vector $ U(N)$ field theory.
The ordinary (commutative) theory can be gauge fixed to the so called unitary
gauge where the standard massive vector field theory is recovered.
Although the same mechanism can be easily extended to the noncommutative theory,
non trivial aspects appear when one considers the Seiberg-Witten map of that theory.
Taking into account the compensating field sector as well as the terms that 
come from the homogeneous equations that define the Seiberg-Witten map, we have found 
several nonequivalent solutions.
One of them consistently admits the implementation of the  unitary gauge 
fixing for all the fields.

\medskip

Acknowledgment: This work is supported in part by CAPES and CNPq (Brazilian
research agencies).

\end{document}